\documentclass{moriond}

\usepackage{amsmath}
\def\be{\begin{equation}}
\def\ee{\end{equation}}
\def\bea{\begin{eqnarray}}
\def\eea{\end{eqnarray}}

\usepackage{xcolor}

\newcommand{\Photo}{}

\begin{document}
\vspace*{4cm}

\title{Cosmic Birefringence from \textit{Planck} Public Release 4}

\author{P. Diego-Palazuelos$^{1,2}$, J. R. Eskilt$^{3}$, Y. Minami$^{4}$, M. Tristram$^{5}$, R. M. Sullivan$^{6}$, A. J. Banday$^{7,8}$, R. B. Barreiro$^{1}$, H. K. Eriksen$^{3}$,  K. M. G\'orski$^{9,10}$, R. Keskitalo$^{11,12}$, E. Komatsu$^{13,14}$, E. Mart\'inez-Gonz\'alez$^{1}$, D. Scott$^{6}$, P. Vielva$^{1}$, and I. K. Wehus$^{3}$}

\address{\scriptsize $^{1}$ Instituto de F\'isica de Cantabria (CSIC-Universidad de Cantabria), Santander, Spain\\
$^{2}$ Departamento de F\'isica Moderna, Universidad de Cantabria, Santander, Spain\\
$^{3}$ Institute of Theoretical Astrophysics, University of Oslo, Oslo, Norway\\
$^{4}$ Research Center for Nuclear Physics, Osaka University, Osaka, Japan\\
$^{5}$ Universit\'e Paris-Saclay, CNRS/IN2P3, IJCLab, Orsay, France\\
$^{6}$ Department of Physics \& Astronomy, University of British Columbia, Vancouver, Canada\\
$^{7}$ Universit\'e de Toulouse, UPS-OMP, IRAP, Toulouse, France\\
$^{8}$ CNRS, IRAP, Toulouse, France\\
$^{9}$ Jet Propulsion Laboratory, California Institute of Technology, Pasadena, U.S.A\\
$^{10}$ Warsaw University Observatory, Warszawa, Poland\\
$^{11}$ Computational Cosmology Center, Lawrence Berkeley National Laboratory, Berkeley, U.S.A.\\
$^{12}$ Space Sciences Laboratory, University of California, Berkeley, U.S.A\\
$^{13}$ Max Planck Institute for Astrophysics, Garching, Germany\\
$^{14}$ Kavli Institute for the Physics and Mathematics of the Universe, Todai Institutes for Advanced Study, University of Tokyo, Kashiwa, Japan}

\maketitle

\abstract{ We search for the signature of parity-violating physics in the Cosmic Microwave Background using \textit{Planck} polarization data from the Public Release 4 (PR4 or \texttt{NPIPE}). For nearly full-sky data, we initially find a birefringence angle $\beta=0.30^\circ\pm0.11^\circ$ ($68\%$~C.L.). We also find that the values of $\beta$ decrease as we enlarge the Galactic mask, which can be interpreted as the effect of polarized foreground emission. We use two independent approaches to model this effect and mitigate its impact on $\beta$. Although results are promising, and the good agreement between both models is encouraging, we do not assign cosmological significance to the measured value of $\beta$ until we improve our knowledge of the foreground polarization. Acknowledging that the miscalibration of polarization angles is not the only instrumental systematic that can create spurious TB and EB correlations, we also perform a detailed study of \texttt{NPIPE} end-to-end simulations to prove that our measurements of $\beta$ are not significantly affected by any of the known systematics. }

\section{Parity-violating physics in the CMB polarization}

To this day, we only fully understand about 5\% of the contents of the Universe, with the remainder of its energy content split into approximately 27\% of Dark Matter (DM), and 68\% of Dark Energy (DE). Numerous models have been proposed to explain these two dark components, \textit{e.g.}~\cite{DM,DE}, a wide range of new weakly interacting massive particles, exotic neutrino models, quintessence, and modified gravity models. Some of them, whether be it as a solution for DM or DE, have in common the introduction of a new pseudoscalar field, $\phi$, that changes sign under the inversion of spatial coordinates, $\phi(-\vec{n})=-\phi(\vec{n})$, thus violating parity conservation. A particularly interesting candidate that predicts this type of pseudoscalar field are axion-like particles~\cite{axions}, which, depending on the value of their mass, could act at the same time as a solution for early DE and then behave like DM at later times.

An interesting property that these parity-violating pseudoscalar fields have in common is that, if coupled to the electromagnetic tensor, $F_{\mu\nu}$, and its dual, $\tilde{F}^{\mu\nu}$, via a Chern-Simons term in the Lagrangian density, $\mathcal{L} \supset \frac{1}{4}g_{\phi\gamma}\phi F_{\mu\nu}\tilde{F}^{\mu\nu}$, they can make the phase velocities of the right- and left-handed helicity states of photons differ~\cite{Carroll1990,Carroll1991,Harari1992}. Such offset has the effect of rotating the plane of linear polarization clockwise on the sky by an angle $\beta=-\frac{1}{2}g_{\phi\gamma}\int \frac{\partial \phi}{\partial t}dt$, where $g_{\phi\gamma}$ is the coupling constant between photons and the new pseudoscalar field. This rotation is what we call ``cosmic birefringence'' because it is as if space itself acted like a birefringent material (see Ref.~\cite{Eiichiro_review} for a review).

Although we know $g_{\phi\gamma}$ must be small, in principle, we could constraint this kind of DM and DE models by measuring the rotation of the plane of polarization of a well-known source of linearly polarized light situated at a far away enough distance as to allow photons to experience a significant $\partial \phi/\partial t$ evolution. Emitted at the epoch of recombination ($z\approx 1100$), and with its polarization angular power spectra accurately predicted by the $\Lambda$ Cold Dark Matter ($\Lambda$CDM) model, the Cosmic Microwave Background (CMB) is, therefore, the perfect tool for the search of cosmic birefringence~\cite{lue}.

CMB polarization can be decomposed into two eigenstates of parity~\cite{zaldarriaga,kamionkowski}: the parity-even E-modes, and the parity-odd B-modes. Expressing these two modes in terms of their spherical harmonic coefficients, $E_{\ell m}^{\mathrm{CMB}}$ and $B_{\ell m}^{\mathrm{CMB}}$, we can calculate their corresponding two-point correlation functions to obtain one parity-odd and two parity-even angular power spectra:
\begin{equation}
\begin{aligned}
        \langle E_{\ell m}^{\mathrm{CMB}}E^{\mathrm{CMB}*}_{\ell'm'}\rangle & = \delta_{mm'}\delta_{\ell\ell'} C_\ell^{EE,\mathrm{CMB}}\\
        \langle B_{\ell m}^{\mathrm{CMB}}B^{\mathrm{CMB}*}_{\ell'm'}\rangle & = \delta_{mm'}\delta_{\ell\ell'} C_\ell^{BB,\mathrm{CMB}}\\
        \langle E_{\ell m}^{\mathrm{CMB}}B^{\mathrm{CMB}*}_{\ell'm'}\rangle & = \delta_{mm'}\delta_{\ell\ell'} C_\ell^{EB,\mathrm{CMB}}
\end{aligned}
\begin{aligned}
&\left.\vphantom{\begin{aligned}
        \langle E_{\ell m}^{\mathrm{CMB}}E^{\mathrm{CMB}*}_{\ell'm'}\rangle & = \delta_{mm'}\delta_{\ell\ell'} C_\ell^{EE,\mathrm{CMB}}\\
        \langle B_{\ell m}^{\mathrm{CMB}}B^{\mathrm{CMB}*}_{\ell'm'}\rangle & = \delta_{mm'}\delta_{\ell\ell'} C_\ell^{BB,\mathrm{CMB}}
  \end{aligned}}\right\rbrace\quad\text{\hspace{-1mm}Parity-even}\\
&\left.\vphantom{\begin{aligned}
        \langle E_{\ell m}^{\mathrm{CMB}}B^{\mathrm{CMB}*}_{\ell'm'}\rangle & = \delta_{mm'}\delta_{\ell\ell'} C_\ell^{EB,\mathrm{CMB}}
  \end{aligned}}\right\rbrace\quad\text{Parity-odd}
\end{aligned}
\end{equation}

In the presence of cosmic birefringence, the intrinsic CMB polarization would then be rotated by an angle $\beta$,
\begin{equation}\label{eq: rotated CMB}
    \begin{pmatrix}
		E_{\ell m}^{\mathrm{o}} \\
	    B_{\ell m}^{\mathrm{o}}
	\end{pmatrix}
	=
	\begin{pmatrix}
		\mathrm{c}(2\beta) & -\mathrm{s}(2\beta)\\
	    \mathrm{s}(2\beta) & \phantom{-}\mathrm{c}(2\beta)
	\end{pmatrix}	
	\begin{pmatrix}
		E_{\ell m}^{\mathrm{CMB}} \\
	    B_{\ell m}^{\mathrm{CMB}}
	\end{pmatrix}\, ,
\end{equation}
so that the observed EB angular power spectra (denoted by the ``o'' superscript) becomes
\begin{equation} \label{eq: observed EB CMB}
    C_\ell^{EB,\mathrm{o}} = \frac{1}{2}\mathrm{s}(4\beta)\Big( C_\ell^{EE,\mathrm{CMB}} - C_\ell^{BB,\mathrm{CMB}}\Big)+ \mathrm{c}(4\beta)C_\ell^{EB,\mathrm{CMB}}.
\end{equation}
For brevity, we refer to the sine, cosine, and tangent functions as $\mathrm{s}$, $\mathrm{c}$, and $\mathrm{t}$. According to $\Lambda$CDM, the Universe has no preferred direction and the statistics of CMB anisotropies should be invariant under parity transformation, yielding a null EB correlation at recombination. In this way, finding $C_\ell^{EB,\mathrm{o}}\neq 0$ would be an evidence of parity-violating physics~\cite{lue}, something that so far has only been observed in the weak interaction. Furthermore, any signal found in the measured EB correlation that resembles that of $C_\ell^{EE,\mathrm{CMB}}$ can be attributed to being the effect of cosmic birefringence.

\section{Joint estimate of birefringence and miscalibrated polarization angles}

Analyses that search for cosmic birefringence in the CMB polarization face two obstacles. Firstly, any miscalibration of the polarization angle of the detector (created, \textit{e.g.}, by a misalignment of the star tracker on the satellite with respect to the telescope mount, or a side effect of the half-wave plate that some future CMB experiments will implement) will also produce the rotation of the plane of linear polarization~\cite{RAC}. For an $\alpha$ miscalibration angle, this means that the observed EB correlation would yield $\beta + \alpha$ instead of $\beta$. And secondly, the CMB is not the only source of polarized emission in the microwave range. Our Galaxy is a bright source of polarized foreground emission, consisting of mainly synchrotron radiation at lower frequencies, and thermal dust emission at higher frequencies.

Fortunately, we can use Galactic foreground emission to break the degeneracy between the $\beta$ and $\alpha$ angles~\cite{Yuto_auto}. Since Galactic foregrounds are produced locally, it is safe to assume that the $\partial \phi /\partial t$ evolution of $\phi$ that they see within our Galaxy is negligible when compared to that experienced by the CMB photons that have been traveling since the epoch of recombination. In this way, Galactic foregrounds would not be significantly affected by cosmic birefringence, and they would only be rotated by the miscalibration of the detector. Updating Eq.~\ref{eq: rotated CMB} to include the foreground signal and a potential $\alpha$ miscalibration, 
\begin{equation}
    \begin{pmatrix}
		E_{\ell m}^{\mathrm{o}} \\
	    B_{\ell m}^{\mathrm{o}}
	\end{pmatrix}
	=
	\begin{pmatrix}
		\mathrm{c}(2\alpha) & -\mathrm{s}(2\alpha) \\
	    \mathrm{s}(2\alpha) & \phantom{-}\mathrm{c}(2\alpha)
	\end{pmatrix}
	\begin{pmatrix}
		E_{\ell m}^{\mathrm{FG}} \\
	    B_{\ell m}^{\mathrm{FG}}
	\end{pmatrix} 
	+
	\begin{pmatrix}
		\mathrm{c}(2\alpha + 2\beta) & -\mathrm{s}(2\alpha + 2\beta)) \\
	    \mathrm{s}(2\alpha + 2\beta) & \phantom{-}\mathrm{c}(2\alpha + 2\beta)
	\end{pmatrix}	
	\begin{pmatrix}
		E_{\ell m}^{\mathrm{CMB}} \\
	    B_{\ell m}^{\mathrm{CMB}}
	\end{pmatrix}\, ,
\end{equation}
it can be proven~\cite{Yuto_cross,my_beta_paper,Johannes} that the observed EB angular power spectrum can be written like
\begin{equation} \label{eq: observed EB CMB + foregrounds}
    C_\ell^{EB,\mathrm{o}} = \frac{\mathrm{t}(4\alpha)}{2}\Big( C_\ell^{EE,\mathrm{o}} - C_\ell^{BB,\mathrm{o}}\Big) + \frac{1}{\mathrm{c}(4\alpha)} C_\ell^{EB,\mathrm{FG}} + \frac{\mathrm{s}(4\beta)}{2\mathrm{c}(4\alpha)} \left( C_\ell^{EE,\mathrm{CMB}} - C_\ell^{BB,\mathrm{CMB}}\right)
\end{equation}
when an intrinsic $C_\ell^{EB,\mathrm{CMB}}=0$ is considered. As initially assumed in Refs.~\cite{Yuto_auto,Yuto_cross,PR3_PRL}, the $C_\ell^{EB,\mathrm{FG}}$ term in Eq.~\ref{eq: observed EB CMB + foregrounds} could be neglected because, according to current experimental constraints~\cite{Planck_dust,Felice_synchrotron}, it is still statistically compatible with being null. From Eq.~\ref{eq: observed EB CMB + foregrounds}, we can build a Gaussian likelihood to simultaneously determine both angles,
\begin{equation}\label{eq: likelihood}
-2 \ln \mathcal{L} = \sum_{\ell=\ell_{\mathrm{min}}}^{\ell_{\mathrm{max}}} \left(\mathbf{A}\bar{C}^\mathrm{o}_\ell-\mathbf{B}\bar{C}^{\mathrm{CMB}}_\ell\right)^T\mathbf{M}_\ell^{-1}\left(\mathbf{A}\bar{C}^\mathrm{o}_\ell-\mathbf{B}\bar{C}^{\mathrm{CMB}}_\ell\right) + \sum_{\ell=\ell_{\mathrm{min}}}^{\ell_{\mathrm{max}}}\ln|\mathbf{M}_\ell|,
\end{equation}
using the information contained in the polarization angular power spectra from the cross-correlation of the different $i,j=1,2,...,N_\nu$ frequency bands of any given CMB experiment, and the theoretical CMB angular power spectra:
\begin{eqnarray}
\bar{C}^\mathrm{o}_\ell =  \begin{pmatrix} C_\ell^{E_iE_j,\mathrm{o}} & C_\ell^{B_iB_j,\mathrm{o}} & C_\ell^{E_iB_j,\mathrm{o}} \end{pmatrix} ^T, &
\bar{C}^\mathrm{CMB}_\ell = \begin{pmatrix} C_\ell^{EE,\mathrm{CMB}} b_\ell^ib_\ell^j\omega_{\ell}^2 & C_\ell^{BB,\mathrm{CMB}}b_\ell^ib_\ell^j\omega_{\ell}^2 \end{pmatrix}^T.
\end{eqnarray}
In this last vector, $b_\ell^i$ and $\omega_{\ell}$ are, respectively, the instrumental beam and pixel window functions. The $\mathbf{M}_\ell$ covariance matrix in Eq.~\ref{eq: likelihood} is $\mathbf{M}_\ell = \mathbf{A} \mathrm{Cov}\left(\bar{C}^\mathrm{o}_\ell, \bar{C}_\ell^{\mathrm{o}T} \right) \mathbf{A}^T$, with $\mathbf{A}$ and $\mathbf{B}$ rotation matrices defined like 
\begin{eqnarray}
\mathbf{A}(\alpha_i,\alpha_j)= \begin{pmatrix} \frac{-\mathrm{s}(4\alpha_j)}{\mathrm{c}(4\alpha_i)+\mathrm{c}(4\alpha_j)} & \frac{\mathrm{s}(4\alpha_i)}{\mathrm{c}(4\alpha_i)+\mathrm{c}(4\alpha_j)} & 1 \\\end{pmatrix}\, , &
\mathbf{B}(\alpha_i,\alpha_j, \beta)= \frac{\mathrm{s}(4\beta)}{2\mathrm{c}(2\alpha_i+2\alpha_j)} \begin{pmatrix} 1 & -1 \\ \end{pmatrix}\,.
\end{eqnarray}
Note that, both the data vector $\bar{C}^\mathrm{o}_\ell$, and the covariance matrix $\mathbf{M}_\ell$, are built from the observed spectra so that the only model needed is that of the CMB angular power spectra in $\bar{C}^\mathrm{CMB}_\ell$. A more detailed description of this methodology is given in Refs.~\cite{Yuto_cross,my_beta_paper,Johannes}. 

This technique was recently applied by Ref.~\cite{PR3_PRL} to polarization data from the 100, 143, 217, and 353 GHz frequency bands of the third data release (PR3) of the \textit{Planck} satellite. Ignoring the potential contribution of the foreground EB correlation, and cross-correlating half-mission splits to reduce the effect of instrumental noise and systematics, they found a birefringence angle of $\beta=0.35^\circ\pm 0.14^\circ$ ($68\%$~C.L.) for nearly the full-sky. Encouraged by this exciting $2.4\sigma$ hint of a signal, we wanted to update this result using data from the latest \textit{Planck} data release~\cite{NPIPE}, known as PR4 or \texttt{NPIPE} reprocessing. In the following sections, we present a summary of the main results and conclusions drawn from that analysis, emphasizing some of the aspects of the study on the impact of instrumental systematics that were not covered in Ref.~\cite{NPIPE_PRL}. All the uncertainties are given at a $68\%$ confidence level (C.L.).

\section{New measurement}

The \texttt{NPIPE} release offers a new reprocessing of the raw, uncalibrated detector data from both the low-frequency (LFI) and high-frequency (HFI) instruments of the \textit{Planck} mission. \texttt{NPIPE} achieves a scale-dependent reduction of the total uncertainty thanks to the addition of the data acquired during repointing maneuvers and a general improvement in the modeling of instrumental noise and systematics. See Ref.~\cite{NPIPE} for a more in-depth description of the \texttt{NPIPE} processing pipeline and its associated data products.

Closely following the analysis done in Ref.~\cite{PR3_PRL}, we work with polarization data from the HFI's 100, 143, 217, and 353 GHz frequency bands. To further reduce the impact of correlated noise and systematics, we work with A/B detector splits and exclude the auto-power spectra of the same maps, \textit{e.g.}, 100A$\times$100A. The cross-correlation of detector splits is more robust against the effects of noise and systematics because detector splits are built from independent subsets of antennas observing at the same frequency, while half-mission splits correspond to different exposure times of all the antennas observing at a common frequency. Being built from different antennas, A/B detector splits can present slightly different miscalibrations. Thus, we must fit a different polarization angle for each of them, $\alpha_i$ with $i=\mathrm{100A},\mathrm{100B},...,\mathrm{353B}$, doubling the number of free parameters with respect to the previous PR3 analysis. As done in Ref.~\cite{PR3_PRL}, we focus on high-$\ell$ data, binning both angular power spectra and covariance matrix from $\ell_{\mathrm{min}}=51$ to $\ell_{\mathrm{max}}=1490$, with $\Delta\ell =20$ spacing ($N_{\mathrm{bins}}=72$), to target the cosmic birefringence angle from the epoch of recombination. We initially ignore the potential contribution of a foreground EB correlation.

For this baseline analysis we find consistent results across four independent groups (PDP, JRE, YM, MT). Each analysis pipeline uses different pseudo-$C_\ell$ estimators (\texttt{PolSpice}~\cite{PolSpice}, \texttt{NaMaster}~\cite{NaMaster}, \texttt{Xpol}~\cite{Xpol}) and implementations: JRE, YM, and MT follow the original implementation~\cite{Yuto_cross,PR3_PRL} and obtain the posterior distribution through Markov chain Monte Carlo methods, while PDP uses an alternative approach in which the maximum likelihood solution is analytically calculated by minimizing the log-likelihood within the small-angle approximation~\cite{my_beta_paper}.

\begin{figure}[h]
    \centering
    \includegraphics[width=14.9cm]{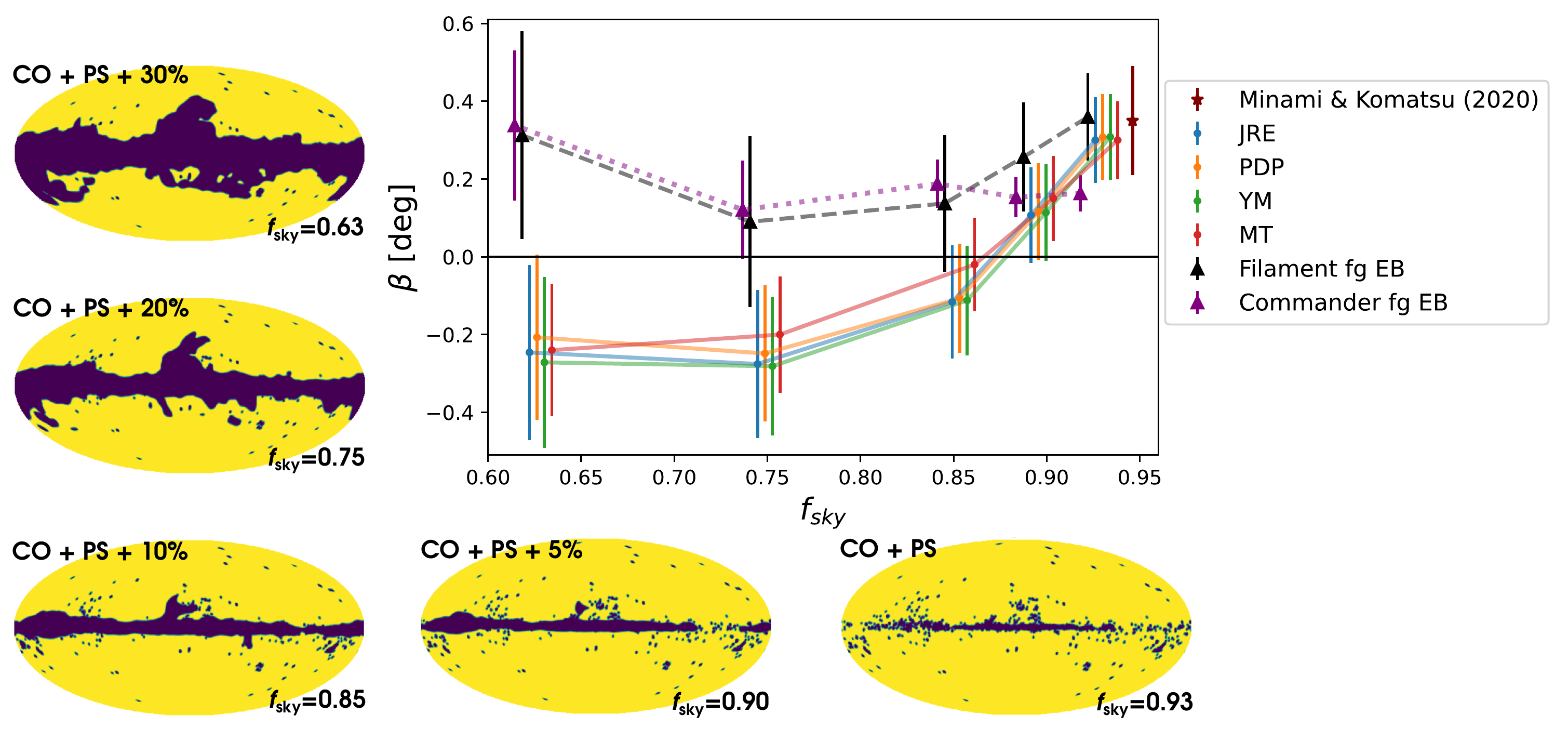}
    \caption{ Constraints on $\beta$ for several sky fractions ($f_{\mathrm{sky}}$) with and without accounting for the foreground EB correlation. For the former, black and purple markers show the corrections using, respectively, the filament model in the JRE pipeline and the \texttt{Commander} sky model in the PDP pipeline. For the latter, the results of four pipelines (JRE, PDP, YM, MT) are shown. The Galactic masks corresponding to each sky fraction are displayed along the left and bottom margins. The birefringence angle obtained for the nearly full-sky analysis of PR3 polarization data is also shown for reference (brown star). However, the $f_{\mathrm{sky}}$ value at which it is plotted is merely illustrative and does not necessarily correspond to the actual effective $f_{\mathrm{sky}}$ from that analysis.}
    \label{fig:beta measurements}
\end{figure}

We start by masking point-like extragalactic sources and pixels where the emission from the carbon monoxide (CO) line is bright (CO+PS mask in Figure \ref{fig:beta measurements}). We mask regions of strong CO emission ($>45 \mathrm{K}_{\mathrm{RJ}}\mathrm{km s}^{-1}$) because, although CO is not polarized, the mismatch of detector bandpasses can create a spurious polarization signal via intensity-to-polarization leakage. As our first result, we obtain a birefringence angle of $\beta=0.30^\circ\pm 0.11^\circ$ for this nearly full-sky configuration. This measurement is compatible with and more precise than the previous result obtained from PR3~\cite{PR3_PRL}.

The birefringence angle that we are looking for is supposed to be an isotropic signal in the sky. Therefore, verifying that we recover compatible values of $\beta$ when masking different regions of the sky would be a good consistency test. However, we found that the measured value of $\beta$ decreased as we started to mask progressively larger regions of the brightest foreground emission in the Galactic plane (5\%, 10\%, 20\%, and 30\% Galactic masks in Figure \ref{fig:beta measurements}). Although it is not shown in Figure \ref{fig:beta measurements}, the decrease on $\beta$ as we enlarge the Galactic mask is accompanied by the complementary increase on $\langle\alpha\rangle = N_\nu^{-1}\sum_i^{N_\nu}\alpha_i$ so that the $\beta + \langle\alpha\rangle$ sum maintains a constant value of $\simeq 0.3^\circ$.

As it was anticipated in Refs.~\cite{Yuto_auto,PR3_PRL,Yuto_cut-sky}, this behavior can be explained by the foreground EB correlation that we have ignored until now. To understand the effect that a non-zero foreground EB correlation might have on our measurements, we can rewrite the observed foreground angular power spectrum to be
\begin{eqnarray}\allowdisplaybreaks
C_\ell^{EB, \mathrm{FG},\mathrm{o}} = & \cfrac{1}{2} \mathrm{s}(4\alpha)\left(C_\ell^{EE, \mathrm{FG}}-C_\ell^{BB,\mathrm{FG}} \right) + \mathrm{c}(4\alpha) C_\ell^{EB, \mathrm{FG}} \allowdisplaybreaks \nonumber \\
 = & \cfrac{1}{2}\sqrt{ 4\left( C_\ell^{EB, \mathrm{FG}}\right)^2 + \left(C_\ell^{EE, \mathrm{FG}}-C_\ell^{BB,\mathrm{FG}} \right)^2}\mathrm{s}(4\alpha+4\gamma_\ell),
\end{eqnarray}
where $\gamma_\ell$ is a new effective angle that, within the small-angle approximation, corresponds to
\begin{equation}
\gamma_\ell \approx \frac{C_\ell^{EB, \mathrm{FG}}}{2\left(C_\ell^{EE, \mathrm{FG}}-C_\ell^{BB,\mathrm{FG}} \right)}.
\end{equation}

From this model, one can see that, if $C_\ell^{EB, \mathrm{FG}}$ is proportional to $C_\ell^{EE, \mathrm{FG}}-C_\ell^{BB,\mathrm{FG}}$, then the $\gamma_\ell$ angle would be independent of the multipole, $\gamma_\ell=\gamma$. In this scenario, $\gamma$ becomes degenerate with $\alpha$, meaning that we will effectively measure $\alpha + \gamma$ instead of $\alpha$, and  $\beta - \gamma$ instead of $\beta$, with the $\alpha +\beta$ sum remaining unaffected~\cite{Yuto_auto}. Previous analyses of \textit{Planck} data have already reported that Galactic dust emission has a positive TB correlation~\cite{Planck_dust}, suggesting that dust could also have a positive EB correlation. A positive $C_\ell^{EB, \mathrm{FG}}$ would give $\gamma>0$, producing a reduction of $\beta$ and an increase of $\alpha$ like the ones seen in our results. In other words, the measured value of $\beta$, which is actually $\beta-\gamma$, is a lower bound for the true value of $\beta$~\cite{PR3_PRL}.

\section{Modeling the impact of the foreground EB correlation}

We now need a model of $C_\ell^{EB, \mathrm{FG}}$ to correct the impact of foregrounds on our measurements and obtain an unbiased estimation of the true underlying birefringence angle. To this end, we use the model for $C_\ell^{EB, \mathrm{FG}}$ proposed in Ref.~\cite{Clark}. In that work, they demonstrate that the misalignment between the filamentary dust structures of the interstellar medium and the plane-of-sky orientation of the Galactic magnetic field induces a non-null EB correlation on Galactic dust emission. If the long axes of filamentary structures are, on average, perfectly aligned with the magnetic field, then null TB and EB correlations are expected. However, a small misalignment between filaments and the magnetic field will produce non-null TB and EB correlations, with their sign depending on the $\psi$ angle of the misalignment. In particular, TB$\propto\sin(2\psi)$, EB$\propto\sin(4\psi)$, and TE$\propto\cos(2\psi)$ correlations are expected.

From their study of the misalignment between the distribution of filaments of neutral hydrogen atoms on synthetic dust simulations and Galactic magnetic field lines derived from \textit{Planck} data, they also conclude that the dust EB correlation produced by this mechanism is strongly dependent on the analysis mask. They expect a small effect for a nearly full-sky configuration, whereas they expect a larger effect when a significant percentage of the Galactic plane is masked. This expectation agrees with the decline in $\beta$ seen on the data as we enlarge the Galactic mask.

With the minor modifications discussed in Refs.~\cite{Johannes,NPIPE_PRL}, we adopt the model proposed by Ref.~\cite{Clark} to predict the amplitude and sign of the dust EB angular power spectrum from the dust EE, TE, and TB correlations:
\begin{equation}\label{eq: clark dust model}
C_\ell^{EB, \mathrm{dust}} \approx 2 A_\ell C_\ell^{EE, \mathrm{dust}} \frac{C_\ell^{TB, \mathrm{dust}}}{C_\ell^{TE, \mathrm{dust}}},
\end{equation}
where we leave $A_\ell$ to be a free amplitude parameter ($0 \leq A_\ell \ll 1$) to fit alongside $\beta$ and $\alpha_i$ in the likelihood. We take the dust EE, TE, and TB angular power spectra from \texttt{NPIPE} data at 353GHz since dust emission dominates over the CMB at that frequency. To account for the possible dependence on $\ell$, we split $A_\ell$ into four bins ($51\leq\ell\leq130$, $131\leq\ell\leq210$, $211\leq\ell\leq510$, and $511\leq\ell\leq1490$). With this correction, we now recover consistent positive values of $\beta$ for all sky fractions (black markers in Figure \ref{fig:beta measurements}). This result confirms our initial hypothesis that the decline in $\beta$ was caused by the foreground EB correlation.

To further corroborate this idea, we also tried another completely independent approach using \texttt{NPIPE} foreground simulations of the \texttt{Commander} sky model. The \texttt{Commander}~\cite{Commander} sky model is built by fitting power-law synchrotron and one-component modified blackbody dust spectral energy distributions (SEDs) to \textit{Planck} data, producing templates of the observed synchrotron and dust emissions in the sky. We used those templates to calculate the angular power spectra of Galactic dust emission and directly introduce them in our equations,
\begin{equation}
C_\ell^{EB,\mathrm{o}} = \frac{\mathrm{t}(4\alpha)}{2} \Big( C_\ell^{EE,\mathrm{o}} - C_\ell^{BB,\mathrm{o}}\Big) + \frac{1}{\mathrm{c}(4\alpha)} \mathcal{D} C_\ell^{EB,\mathrm{dust}} + \frac{ \mathrm{s}(4\beta)}{2\mathrm{c}(4\alpha)} \left( C_\ell^{EE,\mathrm{CMB}} - C_\ell^{BB,\mathrm{CMB}}\right),
\end{equation}
leaving also a free amplitude parameter $\mathcal{D}$ to be simultaneously fitted alongside $\beta$ and $\alpha$ in the likelihood.

This approach also leads to consistent positive values of $\beta$ for all sky fractions (purple markers in Figure \ref{fig:beta measurements}). The results obtained with this model are in very good agreement with the ones from the filament model, except for the nearly full-sky measurement. That discrepancy could merely be a consequence of the difficulties faced by the \texttt{Commander} sky model in capturing the complexity of the dust emission on the center of the Galactic plane.

Another caveat to consider when interpreting the results of this approach is that the \texttt{Commander} SED model does not consider the existence of a potential $\alpha_i$ miscalibration across frequency bands. This might eventually yield a small spurious EB correlation on their foreground maps. In addition, the \texttt{Commander} sky model is derived from previous releases of \textit{Planck} data, and does not yet provide a signal-dominated template for the foreground EB. Thus, \texttt{Commander} foreground maps are very well correlated with \texttt{NPIPE} data, maybe even to the point of reproducing some of its statistical fluctuations. This leads to a reduction of the covariance matrix and the smaller uncertainties seen in Figure \ref{fig:beta measurements}.

On the other hand, the filament model might be limited in its simplicity, and not fully capture the complex interplay of dust and magnetic field lines at all angular scales. In that sense, these two approaches are complementary, and, although it is encouraging that they yield similar results, we need to improve our understanding of the foreground EB before achieving a definitive measurement of $\beta$.

\section{Quantifying systematics using the end-to-end simulations}

The miscalibration of the polarization angle of the detector is not the only instrumental effect that can potentially affect our analysis. Systematic effects like intensity-to-polarization leakage, beam leakage, or cross-polarization effects, can also produce spurious TB and EB correlations. To assess the impact of such systematics on our measurements, we conducted a detailed study of the official \texttt{NPIPE} end-to-end simulations. Although briefly commented in Ref.~\cite{NPIPE_PRL}, further results of that study will be presented in Ref.~\cite{my_beta_paper}.

\texttt{NPIPE} end-to-end simulations are built by passing the expected CMB and foreground signals for each frequency band through the full \textit{Planck} instrument model and the \texttt{NPIPE} processing pipeline. In addition to the CMB and foreground signals, the frequency maps produced in this way also capture the instrumental noise and systematics, and the non-linear response of the instrument that eventually leads to couplings between noise and signal. ``Residuals'' maps, which contain only instrumental noise and systematics, are then produced by subtracting the initial input CMB and foreground signals from those frequency maps. See Ref.~\cite{NPIPE} for a more technical description of \texttt{NPIPE} end-to-end simulations. 
 
Since the effect of Galactic foregrounds was already determined, here we focus on the effect of systematics by using simulations of just CMB and Residuals. However, without foregrounds, we are no longer able to break the degeneracy between the birefringence and polarization angles. Therefore, we can either fit a different angle for each detector split, \textit{i.e.}, $\alpha_i$, or we can fit the same angle for all frequency bands, \textit{i.e.}, $\beta$. 

\begin{figure}[h]
    \centering
    \includegraphics[width=8.5cm]{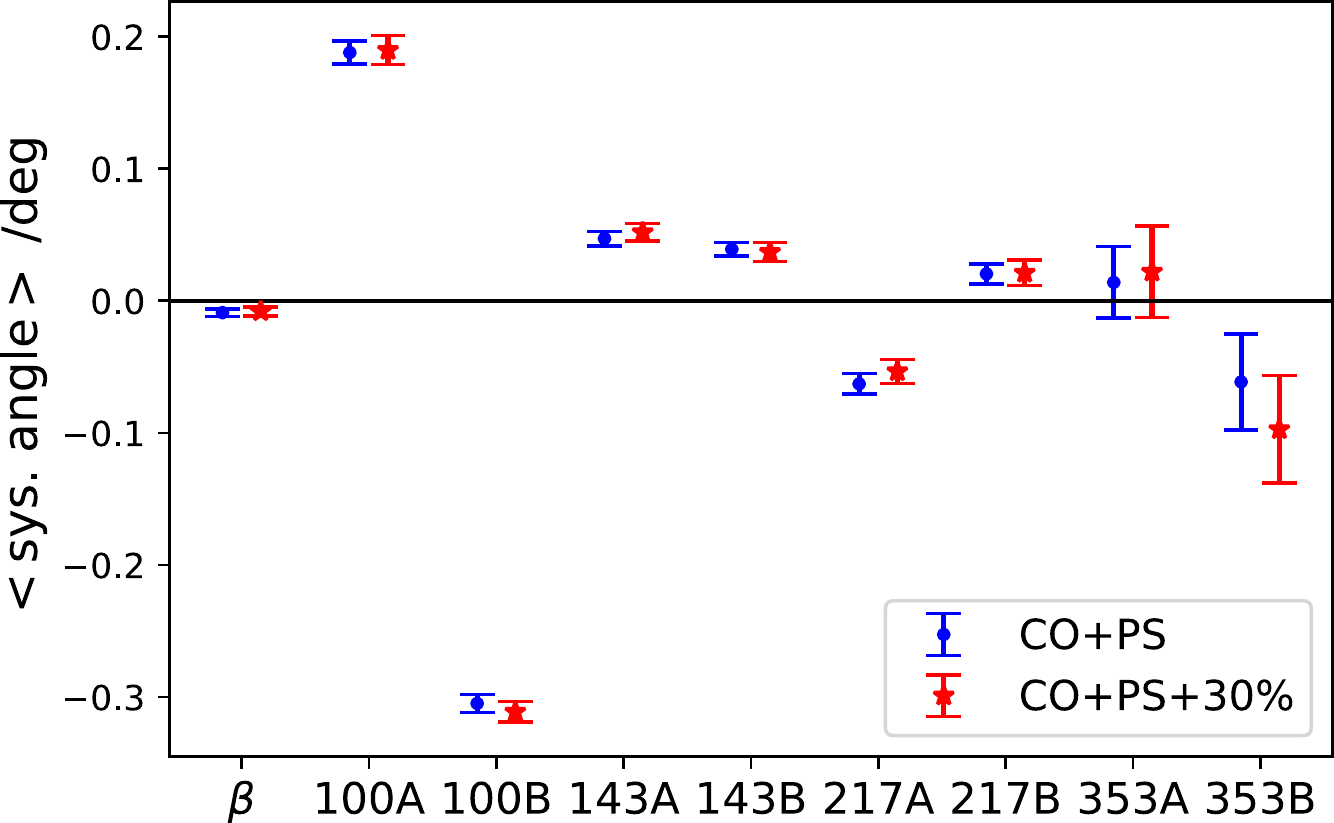}
    \caption{Mean systematic birefringence and polarization angles found in \texttt{NPIPE} CMB + Residuals simulations. Angles are averaged over $N_{\mathrm{sim}}=100$ simulations, with uncertainties calculated as the simulations' dispersion over $\sqrt{N_{\mathrm{sim}}}$. The blue points show the results for our smallest mask (CO+PS, $f_{\mathrm{sky}}=0.93$), while the red points are for our largest mask (CO+PS+30\%, $f_{\mathrm{sky}}=0.63$).}
    \label{fig:systematic angles}
\end{figure}

For this test, we built a set of 100 simulations by coadding each CMB realization with its associated Residual map, which we mask and analyze as we did with the data. Taking turns fitting angles that behave either like $\alpha_i$ or $\beta$, we obtain the mean systematic angles shown in Figure \ref{fig:systematic angles}. We find the presence of some systematic $\alpha_i$ angles on the simulations, especially for the 100A and 100B detector splits. To understand their origin, we performed a closer study of the simulations, finding that the angular power spectra of CMB + Residuals simulations at the 100A and 100B frequency bands resemble that of $C_\ell^{EE,\mathrm{CMB}}$. In general, intensity-to-polarization leakage gives $C_\ell^{EB}\propto C_\ell^{TT}$, whereas the cross-polarization effect gives $C_\ell^{EB}\propto C_\ell^{EE}$, and a combination of the two would give $C_\ell^{EB}\propto C_\ell^{TE}$. Thus, we believe that the systematic angles found in the simulations are due to a cross-polarization effect. This kind of systematic is particularly dangerous for our analysis since our estimator relies precisely on finding a $C_\ell^{EE,\mathrm{CMB}}$-like signal in the measured EB correlation to determine both the birefringence and polarization angles.

Identifying the existence of this cross-polarization effect is essential to understand all the effects at play in both simulations and data. Nevertheless, note that the systematic angles found on the simulations do not need to agree with the ones found in the data because the simulations do not, and in fact cannot, include the actual unknown miscalibration angles present in the data. In this way, the main conclusion to draw from these results is that, even in the presence of such cross-polarization effect, and the rest of the known systematics, our methodology is able to correctly capture their effect within the $\alpha_i$ miscalibration angles, leaving the measurement of $\beta$ not significantly affected by any of them. This observation justifies our decision not to correct our $\beta$ measurements for any of the known systematics.
 
Another conclusion to draw from the results in Figure 2 is that we find consistent mean systematic angles for our largest (CO+PS+30\%) and smallest (CO+PS) masks. This means that none of the known systematics can reproduce the decline in $\beta$ as we enlarge the Galactic mask, reinforcing our hypothesis that it is driven by the foreground EB correlation.

\section{Conclusions}

In this work, we continue the search for cosmic birefringence in the CMB polarization by updating the previous analysis of \textit{Planck} PR3 data~\cite{PR3_PRL} with the latest \texttt{NPIPE} data release. We initially find a birefringence angle of $\beta=0.30^\circ \pm 0.11^\circ$ for nearly full-sky data, which is consistent with and more precise than the previously reported value for PR3.
Exploring the dependence of $\beta$ on Galactic masks, we found that our measurements of birefringence decreased as we enlarged the Galactic mask, which can be interpreted as the effect of the EB correlation of polarized dust emission. We have used two independent models to account for this component, and although results are promising and the good agreement between both models is encouraging, we chose not to assign cosmological significance to our measurement of $\beta$ until we have a better understanding of polarized foreground emission.

If confirmed, cosmic birefringence would be an evidence of physics beyond the standard model of cosmology and particle physics. To make progress, we need to continue the search in independent datasets, especially those with access to the full-sky, like the LiteBIRD~\cite{LiteBIRD} mission. A first follow-up analysis incorporating \textit{Planck} LFI data and exploring the frequency-dependence of the birefringence signal has already been published~\cite{Johannes}. We also need to improve our knowledge of the EB science, both in the sense of achieving a better understanding of the foreground emission, and a better control of the systematics that plague this channel. In this regard, we wanted to stress the need for high-fidelity end-to-end simulations, which have played a paramount role in this analysis by helping to understand all of the systematics affecting the EB correlation.

Last but not least, we can avoid Galactic foregrounds altogether if we do not rely on the measured EB correlation for calibration. To this end, we must improve upon the accuracy of calibrating the artificial rotation of polarization angles due to telescopes, optics, and detectors. Our result suggests that the target accuracy should be well below $0.1^\circ$, \textit{e.g.}, $0.06^\circ$ for a $5\sigma$ result for $\beta\simeq 0.3^\circ$. While challenging, the current technology should allow for such precision~\cite{calibration_satellite_satellite,calibration_Earth_balloon,calibration_Earth_ground}. This line of research should be pursued to obtain the most robust measurement of cosmic birefringence.

\section*{Acknowledgments}

PDP thanks the organizers for the opportunity to present this work at the Cosmology session of the 56$^{\mathrm{th}}$ \textit{Rencontres de Moriond} and for such a great conference. PDP thanks the Spanish Agencia Estatal de Investigación (AEI, MICIU) for the financial support provided under the project with reference PID2019-110610RB-C21, and also acknowledges funding from the Formaci\'on del Profesorado Universitario (FPU) program of the Spanish Ministerio de Ciencia, Innovaci\'on y Universidades, and the Unidad de Excelencia María de Maeztu (MDM-2017-0765). The results presented here are based on observations obtained with \textit{Planck}, an ESA science mission with instruments and contributions directly funded by ESA Member States, NASA, and Canada. This research used resources of the National Energy Research Scientific Computing Center (NERSC), a U.S. Department of Energy Office of Science User Facility operated under Contract No. DE-AC02-05CH11231. We acknowledge the use of \texttt{HEALPix}~\cite{healpix}, \texttt{PolSpice}~\cite{PolSpice}, \texttt{NaMaster}~\cite{NaMaster}, \texttt{Xpol}~\cite{Xpol}, \texttt{emcee}~\cite{emcee}, \texttt{Matplotlib}~\cite{matplotlib}, and \texttt{NumPy}~\cite{numpy}.


\section*{References}

\begin{thebibliography}{99}

\bibitem{DM} J. L. Feng, Annu. Rev. Astron. Astrophys., \textbf{48} (2010). 

\bibitem{DE} J. Yoo and Y. Watanabe, Int. J. Mod. Phys. D, \textbf{21}, 1230002 (2012). 

\bibitem{axions} D. J. E. Marsh, Phys. Rep., \textbf{643} (2016). 

\bibitem{Carroll1990} S. M. Carroll \textit{et al.}, Phys .Rev. D, \textbf{41}, 1231 (1990). 

\bibitem{Carroll1991} S. M. Carroll and G. B. Field, Phys. Rev. D, \textbf{43}, 3789 (1991). 

\bibitem{Harari1992} D. Harari and P. Sikivie, Phys. Lett. B, \textbf{289}, 67 (1992). 

\bibitem{Eiichiro_review} E. Komatsu, arXiv:2202.13919 (2022). 

\bibitem{lue} A. Lue \textit{et al.}, Phys. Rev. Lett., \textbf{83}, 1506 (1999). 

\bibitem{zaldarriaga} M. Zaldarriaga and U. Seljak, Phys. Rev. D, \textbf{55}, 1830 (1997). 

\bibitem{kamionkowski} M. Kamionkowski \textit{et al.}, Phys. Rev. D, \textbf{55}, 7368 (1997). 

\bibitem{RAC} N. Krachmalnicoff \textit{et al.}, J. Cosmol. Astropart. Phys., \textbf{2022}, 039 (2022). 

\bibitem{Yuto_auto} Y. Minami \textit{et al.}, Prog. Theor. Exp. Phys., \textbf{2019}, 083E02, (2019). 

\bibitem{Yuto_cross} Y. Minami and E. Komatsu, Prog. Theor. Exp. Phys., \textbf{2020}, 103E02 (2020). 

\bibitem{my_beta_paper} P. Diego-Palazuelos \textit{et al.}, Manuscript in preparation (2022). 

\bibitem{Johannes} J. R. Eskilt, arXiv:2201.13347 (2022). 

\bibitem{PR3_PRL} Y. Minami and E. Komatsu, Phys. Rev. Lett., \textbf{125}, 221301 (2020). 

\bibitem{Planck_dust} Planck Collaboration, Y. Akrami \textit{et al.}, Astron. Astrophys., \textbf{641}, A11 (2020). 

\bibitem{Felice_synchrotron} F. A. Martire \textit{et al.}, Accept. J. Cosmol. Astropart. Phys., arXiv:2110.12803 (2022). 

\bibitem{NPIPE} Planck Collaboration, Y. Akrami \textit{et al.}, Astron. Astrophys., \textbf{643}, A42 (2020). 

\bibitem{NPIPE_PRL} P. Diego-Palazuelos \textit{et al.}, Phys. Rev. Lett., \textbf{128}, 091302 (2022). 
 
\bibitem{PolSpice} G. Chon \textit{et al.}, Mon. Not. Roy. Astron. Soc., \textbf{350}, 914 (2004). 

\bibitem{NaMaster} D. Alonso \textit{et al.}, Mon. Not. Roy. Astron. Soc., \textbf{484}, 4127 (2019). 

\bibitem{Xpol} M. Tristram \textit{et al.}, Mon. Not. Roy. Astron. Soc., \textbf{358}, 833 (2005). 

\bibitem{Yuto_cut-sky} Y. Minami, Prog. Theor. Exp. Phys., \textbf{2020}, 063E01, (2020). 

\bibitem{Clark} S. E. Clark \textit{et al.}, Astrophys. J., \textbf{919}, 53 (2021). 

\bibitem{Commander} Planck Collaboration, Y. Akrami \textit{et al.}, Astron. Astrophys., \textbf{641}, A4 (2020). 

\bibitem{LiteBIRD} LiteBIRD Collaboration, E. Allys \textit{et al.}, arXiv:2202.02773 (2022). 
 
\bibitem{calibration_satellite_satellite} F. J. Casas \textit{et al.}, Sensors \textbf{2021}, 21(10), 3361 (2021). 

\bibitem{calibration_Earth_balloon} F. Nati \textit{et al.}, J. Astron. Instrum., \textbf{6}, 2, 1740008 (2017). 

\bibitem{calibration_Earth_ground} M. F. Navaroli \textit{et al.}, Proceedings of the SPIE, \textbf{10708}, 107082A (2018). 

\bibitem{healpix} K. M. G\'orski \textit{et al}.,  Astrophys. J., \textbf{622}, 759 (2005). 

\bibitem{emcee} D. Foreman-Mackey \textit{et al}., Publ. Astron. Soc. Pac., \textbf{125}, 925, 306 (2013). 

\bibitem{matplotlib} J. D. Hunter, Comput. Sci. Eng., \textbf{9}, 3 (2007). 

\bibitem{numpy} C. R. Harris \textit{et al}., Nature, \textbf{585}, 7825 (2020). 


\end{thebibliography}
\end{document}